\documentclass[pdflatex,sn-mathphys-num]{sn-jnl}

\usepackage{graphicx}%
\usepackage{multirow}%
\usepackage{amsmath,amssymb,amsfonts}%
\usepackage{amsthm}%
\usepackage{mathrsfs}%
\usepackage[title]{appendix}%
\usepackage{xcolor}%
\usepackage{textcomp}%
\usepackage{manyfoot}%
\usepackage{booktabs}%
\usepackage{url}
\usepackage{algorithm}%
\usepackage{algorithmicx}%
\usepackage{algpseudocode}%
\usepackage{listings}%
\usepackage{tikz}
\usepackage{graphicx}
\usepackage{float}
\usepackage{amsmath}
\usepackage{hyperref}

\theoremstyle{thmstyleone}%
\theoremstyle{thmstyletwo}%

\theoremstyle{thmstylethree}%

\usepackage{xcolor}
\usepackage{upgreek}

\usepackage{etoolbox}   
\makeatletter
\patchcmd{\@maketitle}{\artauthors}{{\artauthors}}{}{}
\makeatother

\begin{document}

\title[Article Title]{Graph representations of dislocation networks}


\author[1]{\fnm{Gauravkumar} \sur{Lathiya}}\email{gauravkumar.lathiya@rub.de}

\author*[1]{\fnm{Markus} \sur{Stricker}}\email{markus.stricker@rub.de}

\affil[1]{\orgdiv{Interdisciplinary Centre for Advanced Materials Simulation(ICAMS)}, \orgname{Ruhr-Universit\"at Bochum }, \orgaddress{\street{Universit\"atstraße 150}, \city{Bochum}, \postcode{44801}, \state{NRW}, \country{Bochum}}}


\abstract{The mechanical response of a crystal is governed by the evolving network of dislocations it contains, yet quantitatively linking the `state of that network' to its yield stress remains challenging. Here we represent the dislocation networks obtained from discrete dislocation dynamic simulation as graphs, using five distinct graph representations, characterize each network through a compact set of topological properties, and predict the yield stress from these graph properties and the applied strain rate using a standard random forest model evaluated with strain rate stratified cross-validation. The model predicts the yield stress with $R^2\approx0.86-0.90$ across all box sizes and $R^2\approx0.91-0.94$ for larger box sizes, and this accuracy is consistent across all five representations, indicating that the topology-stress relationship is a robust feature of the graph properties rather than an artefact  of any single graph representation. We propose graph topology as a compact and robust route for connecting the microstructural state of a dislocation network to its mechanical response in addition to established variables.}


\maketitle

\section{Introduction}\label{sec1}

Dislocations, which represent deviations or irregularities in the crystal structure, defy the behavior expected of point-like particles. Instead, they exhibit behavior similar to the twigs of a bird’s nest~\cite{cottrell1954dislocations}.
From their initial proposal~\cite{Taylor} as the carriers of plastic deformation, dislocations have been the subject of extensive study, with particular focus on how their structures evolve under different conditions and influence macroscopic material properties.
In many safety-critical situations, the mechanical behavior of metals during service is strongly influenced by the rate at which loads are applied.
This is particularly relevant in cases such as vehicle crashes, metal forming processes, medical implants, or bird strikes on jet engines. Generally, metals that deform through dislocation slip~\cite{LINDHOLM1964317} exhibit strain rate hardening.
However, under a limited range of conditions, interactions involving dislocations can be lead to strain rate softening~\cite{ZAISER1997, Zaiser1999}. Understanding how dislocation structures evolve and interact under different loading conditions is therefore essential for establishing a connection between their underlying mechanisms and the resulting macroscopic mechanical response, and ultimately for developing physics-based material models.

Consequently, considerable effort has been devoted to characterizing dislocation structures across different length scales. At the atomic scale, the current understanding of dislocations is largely based on the properties of individual dislocations, particularly their core structure and interactions, as obtained from atomistic simulations~\cite{Vaild, ZepedaRuiz2021Atomistic}.
However, such approaches are inherently limited by computational constraints and the accuracy of interatomic potentials, restricting their applicability to small length and time scales~\cite{Deng, ZepedaRuiz2017Probing}.
Furthermore, the description of extended defect structures requires system sizes that are inaccessible to first-principles methods~\cite{Aleksei}.
Consequently, a fundamental gap remains in linking individual dislocation properties to their collective behavior governing the macroscopic material response.
Another approach to understand dislocation behavior can be based on continuum dislocation dynamics theory (CDD)~\cite{HOCHRAINER2014167, LIN2021102943}.

CDD provides a continuum description of plastic deformation by tracking the evolution and transport of the dislocation density field. In this framework, the evolution of the dislocation density is governed by dislocation fluxes, which in turn determine the associated plastic deformation~\cite{Ebrahimi2014}.
However, the approach is subject to several limitations, including the reliance on closure approximations, the restricted ability to capture inherently discrete dislocation interactions, and the challenges associated with computational complexity and experimental validation~\cite{SANDFELD2015, HOCHRAINER2016}.
A density field approach~\cite{Udofia2025} can be used to characterize complex dislocation systems and provides a methodology for quantitatively comparing dislocation structures.
This approach enables the construction of dislocation structure maps and, consequently, defines a ``coordinate system'' for dislocation configurations.
Embedding maps can offer insight into the dynamic nature of dislocation structures, their relationships, and provide a means to characterize and compare their evolution.
However, the sensitivity of such embeddings to (hyper)parameters suggests that representing dislocations solely through density field descriptors alone may not provide a sufficiently robust metrics for dislocation structures. These limitations motivate the search for descriptors that reflect microstructure can be highly complex even though collective dislocation dynamics may still be restricted by common kinetic relationships, such that accessible dislocation states occupy only a limited region of a general state space.
 
\begin{figure}[H]
    \centering
    \begin{minipage}{0.49\textwidth}
        \centering
        \includegraphics[width=\linewidth]{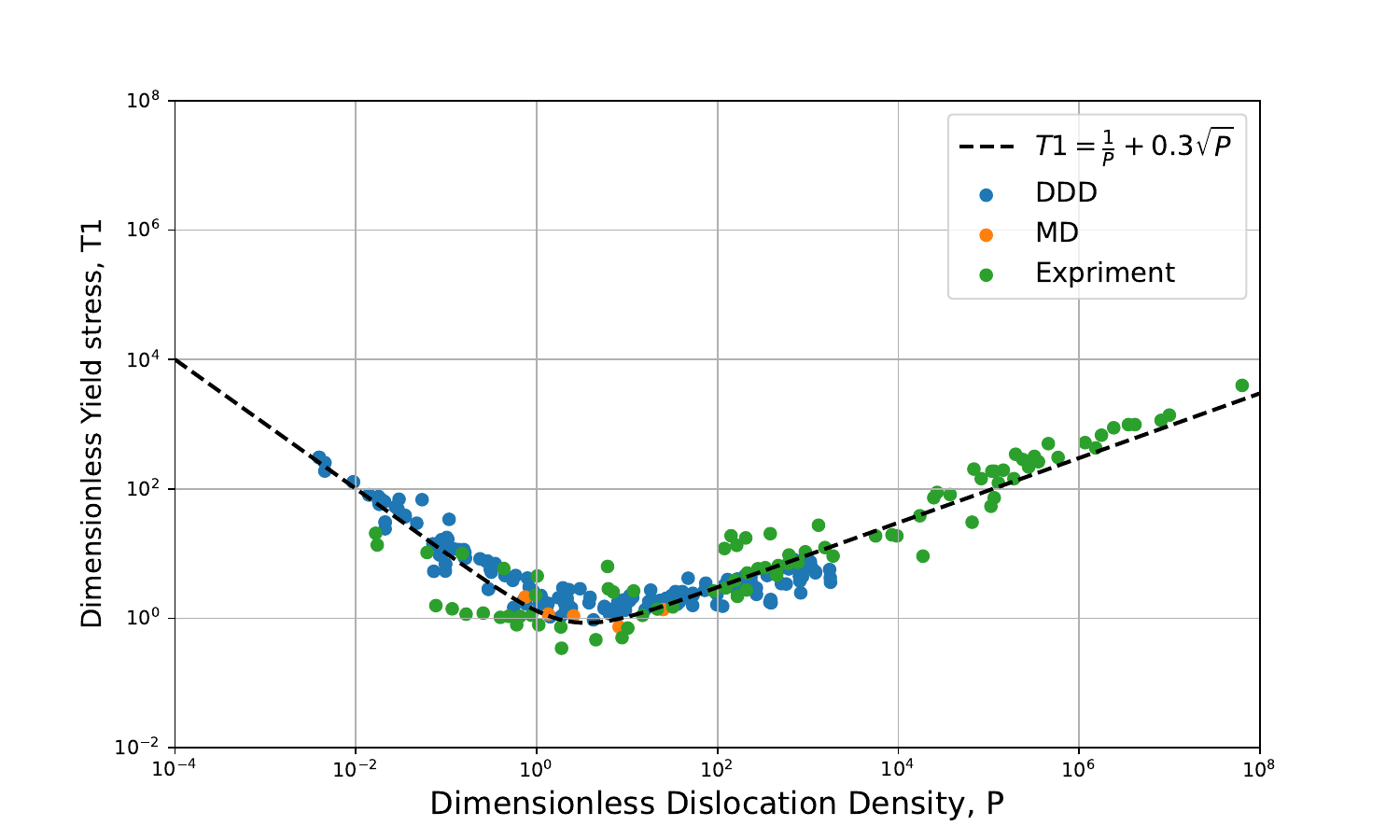}
    \end{minipage}
    \hfill
    \begin{minipage}{0.49\textwidth}
        \centering
        \includegraphics[width=\linewidth]{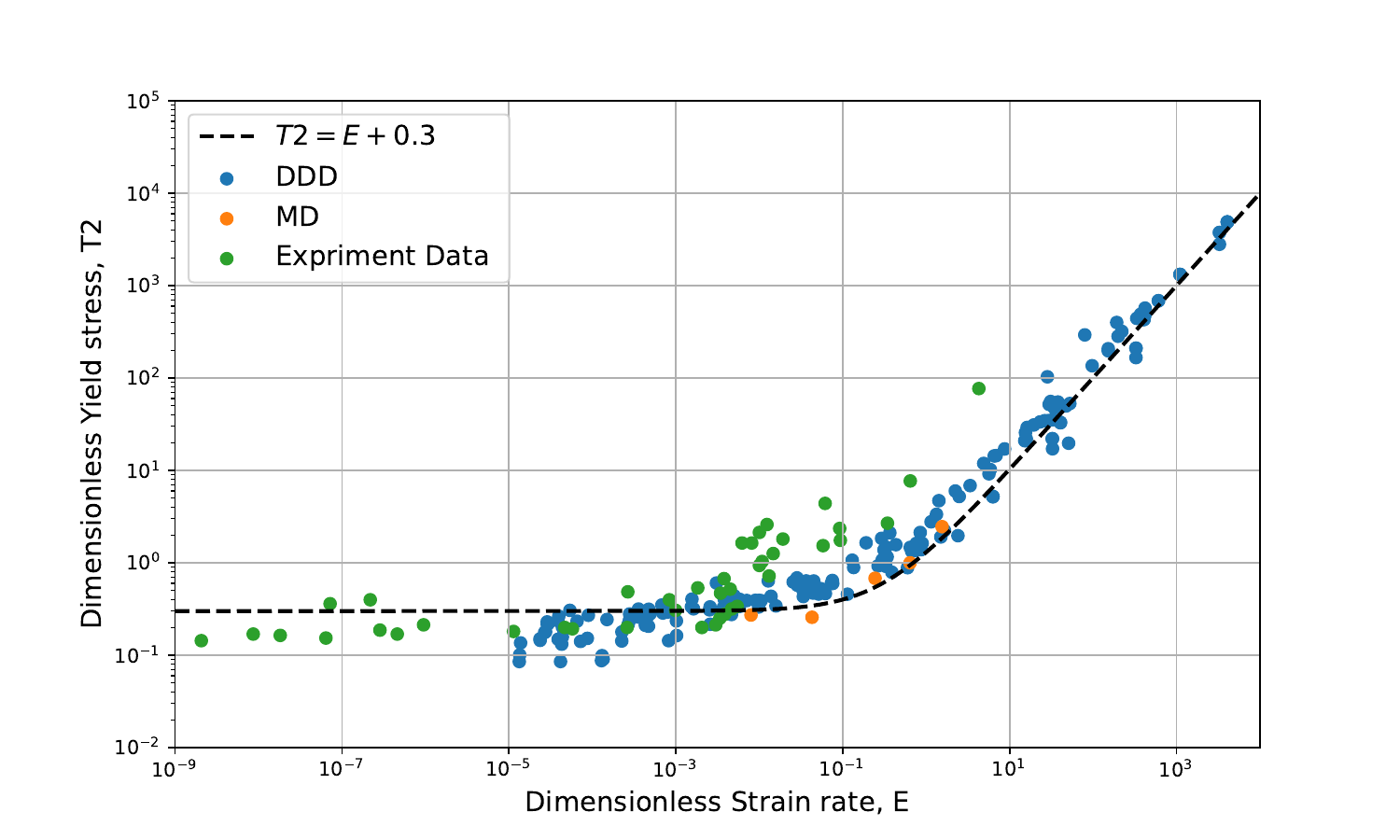}
    \end{minipage}
    
    \caption{Left side plot is dimensionless yield stress versus dimensionless dislocation density; right side plot shows dimensionless yield stress versus dimensionless strain rate. DDD is abbreviation of discrete dislocation dynamics and MD is molecular dynamics~\cite{Fan2021}.}
    \label{fig:fan_P_E_T}
\end{figure}

Evidence for such constrained behavior can be found in the relationship between material strength on strain rate and dislocation density which has been studied through both simulations and experiments. Based on these observations, Fan et al.~\cite{Fan2021} showed that, despite the difference in material systems and deformation conditions, the plastic deformation behavior of various materials can be described within a common normalized framework when appropriately normalized.
To capture this behavior, they proposed a dislocation kinetic model relating normalized strength to normalized strain rate and dislocation density through dislocation mobility.
The model shows good agreement with both simulation and experimental data over a wide range of strain rates, dislocation densities, and material systems.
Their study reveals a transition in strengthening behavior Figure~\ref{fig:fan_P_E_T}: with increasing dislocation density, the material first exhibits a decreasing strength regime associated with strain rate hardening, followed by an increasing strength regime governed by classical forest hardening. Experimental investigations on single crystalline Cu, Al, and LiF, as well as polycrystalline Cu and Al, spanning several orders of magnitude in strain rate, demonstrate that the yield stress is only weakly strain rate-dependent at low strain rates but shows pronounced strain rate hardening at higher strain rates.
The strain rate-independent regime is commonly attributed to dislocation forest interactions and interactions with microstructural obstacles such as grain boundaries or precipitates. 

The collapse of different material systems and deformation conditions onto a common \textit{manifold} suggests that dislocation-mediated deformation is governed by underlying commonalities despite the complexity of individual dislocation structures.
However, while quantities such as the strain rate and dislocation density provide an effective description of the resulting mechanical response, they represent a simplification and do not contain sufficient information to characterize the underlying dislocation state itself.
In particular, distinct dislocation configurations may exhibit similar values of these scalar quantities while differing substantially in their spatial organization and interactions.
This distinction motivates examining whether conventional variables used to describe dislocation-mediated plasticity are also the appropriate variables for characterizing dislocation structures and their evolution.

While conventional variables provide metrics for the collective state of a dislocation population, they do not, in general, retain sufficient information to resolve the underlying spatial organization and connectivity of dislocation network.
A more complete characterization requires descriptors that incorporate structural information beyond reduced measures. Accordingly, several studies have investigated higher order representations aimed at capturing these additional features. In this regard, the concept of an alignment tensor~\cite{Weger2021} and graph based approaches provide promising routes toward a more comprehensive characterization of collective behavior of dislocation structures.
Graph based representations seem well suited to dislocation network because they can explicitly encode connectivity and network topology. 

Previous studies have demonstrated the potential of such approaches. For instance, Starkey et al.~\cite{STARKEY2022} introduced the junction point density as a variable for characterizing the spatial distribution of junction within crystal structures. Their work employed graph-theoretical concepts to analyze network connectivity and establish relationships between the endpoints of open dislocation lines.
Similarly, Katzer et al.~\cite{KATZER2024115841} utilized graph database to analyze three-dimensional dislocation networks obtained from DDD simulations. Their framework enables the efficient extraction of subgraphs and associated structural features, thereby providing a richer description of the evolution of dislocation networks.

Collectively, these studies establish network topology and connectivity as potentially informative characteristics of dislocation structures, provide a basis for investigating whether the collective evolution of dislocation networks can be characterized through their graph representation. 

Building on this premise, our present study investigates the collective behavior of dislocations through graph presentations of the dislocation networks.
In our framework, dislocation structures are mapped onto graphs using different approaches.
The underlying data is obtained from discrete dislocation dynamics (DDD) simulations.
DDD simulations~\cite{kubin_1992, Van_der_Giessen_1995,  Ghoniem_2000, WEYGAND2005, D_Weygand_2002, ZBIB20021133,  Arsenlis_2007} provide a framework for studying dislocation mediated plasticity by resolving the trajectories and interaction of individual dislocations while retaining the emergent collective behavior of the evolving dislocation network.
In these simulations, dislocations are modeled as discrete elastic lines, and essential mechanisms such as glide, cross slip, multiplication~\cite{STRICKER2018319, SUDMANNS2019, Akhondzadeh2021SlipFree}, annihilation, long range interactions, and junction formation are incorporated.
By resolving the geometry, motion, and interaction of individual dislocation throughout deformation, DDD provides direct access to evolving network structure and thus enables relationships between microscopic dislocation organization and macroscopic mechanical responded to be investigated~\cite{Fan2021}.
This explicit network information can, in turn, be represented in graph form, providing a framework for quantifying the topology and connectivity~\cite{AKHONDZADEH2026connectivity, BERTIN2024119884} of dislocation structures.
This paper is organized as follows. Section~\ref{subsec:graph_R} presents the different approaches for representing dislocation networks within a graph theoretical framework. Subsequently, graph based properties are quantified using established Python libraries, namely \texttt{NetworkX}~\cite{NetworkX} and \texttt{NetworKit}~\cite{staudt2015networkit}, enabling a systematic characterization of the network topology.

The resulting feature sets correspond to high dimensional representations of the dislocation networks. For analysis and interpretation, dimensionality reduction technique is applied to project these high dimensional properties into a lower dimensional embedding space. The resulting embeddings are then visualized and analyzed, with results and discussion provided in Sections~\ref{sec:results} and~\ref{sec:discussion}, respectively.

\section{Methods}\label{Methods}

Using discrete dislocation dynamics (DDD), we generate a dataset comprising 627 dislocation structure trajectories from single crystal tensile simulations with the Karlsruhe DDD code~\cite{D_Weygand_2002, WEYGAND2005}.
We vary the applied strain rate, initial dislocation structure, size, and crystallographic orientations across the data set.
The DDD framework provides direct access to the loop based topology of the evolving dislocation network and allows us to tract individual dislocation loops through the deformation process.
We record snapshots of the dislocation configuration at frequent intervals to capture the temporal evolution of the microstructure.

We subsequently construct graph representations of the resulting dislocation networks using several approaches, which we describe in detail in Section~\ref{subsec:graph_R}.

\subsection{Graph representations}\label{subsec:graph_R}
Graph theory offers an abstract formalism for describing relationships (edges) among entities (nodes). Dislocation networks obtained from DDD naturally lend themselves to graph representation, as their computational discretizations already consists of vertices connected by dislocation segments.
However, the mapping of physical components of the dislocation network onto graph nodes and edges is not uniquely defined.
Different interpretations are admissible depending on the structural information of interest, as demonstrated, for example, by graph representations proposed by Starkey et al.~\cite{STARKEY2022} and Katzer et al.~\cite{KATZER2024115841}.

In this section, we denote the different graph constructions as Representation 0 through Representation 4, or short $R_0$-$R_4$.
These representations differ in the level of structural information retained and in the way dislocation features are mapped onto graph nodes and edges.
Starting from the segment level description in $R_0$, subsequent representations progressively abstract or reorganize the dislocation network to emphasize different aspects of its connectivity and topology.
Thus, from $R_0$ to $R_4$, the graph construction systematically changes while preserving \textit{some} features of the original underlying dislocation network.

\subsubsection*{Representation 0 ($R_0$)}\label{subsubsec:R0}
\begin{figure}[H]
\centering 
 \includegraphics[scale=0.4]{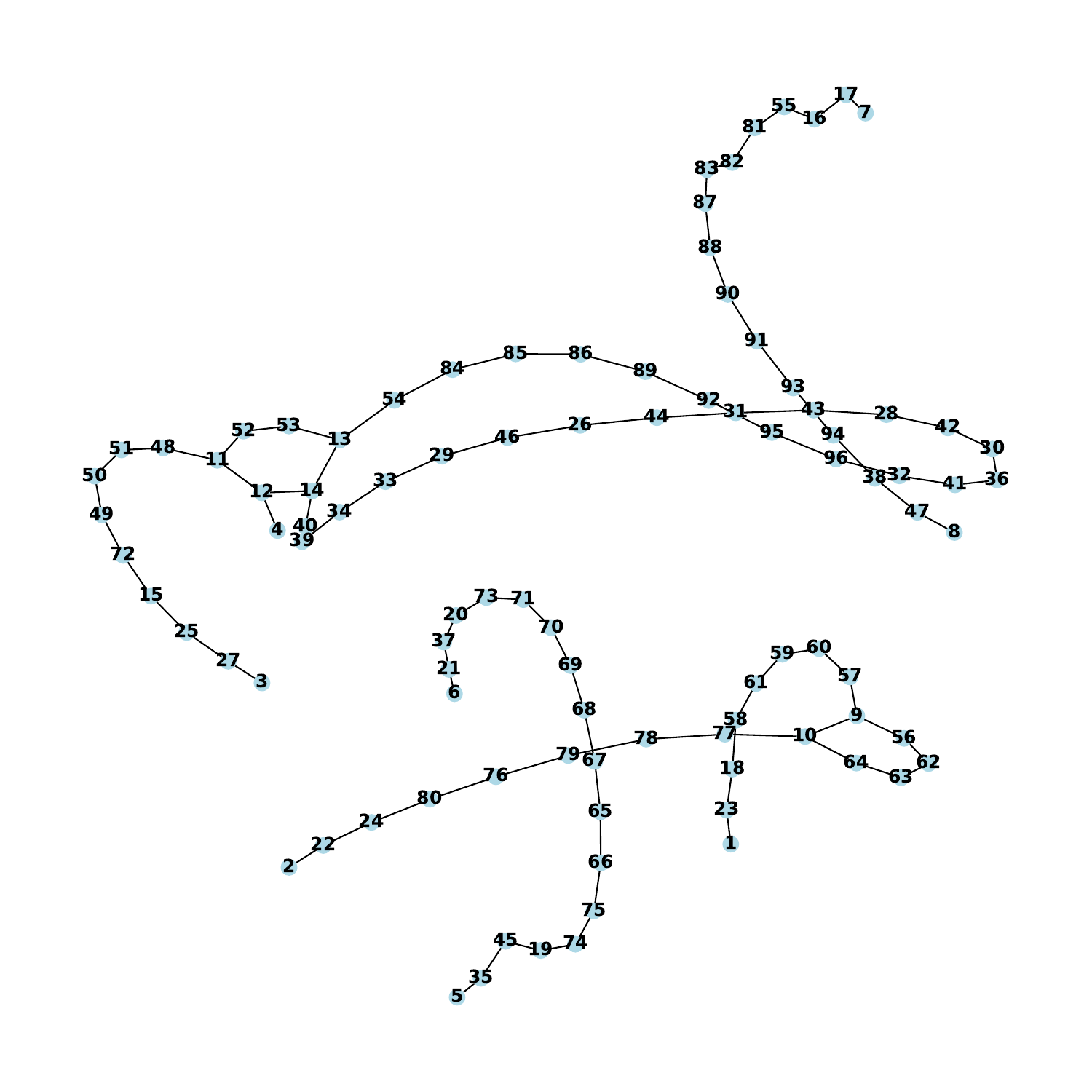}
\caption{For $R_0$, the computational vertices and segments are represented as graph nodes and edges, respectively. The trajectories correspond to discretized dislocation loops. The edges, $(11,12)$, $(13,14)$ and $(10,9)$, represent junctions in the dislocation network.} 
\label{fig:R0}
\end{figure}

The most straightforward approach to representing a dislocation structure as a graph is to directly use the native output of the DDD simulations, the computational vertices and connecting segments.
By interpreting vertices as nodes and segments as edges, the original topology can be preserved without additional transformation.
Note: while the topology is preserved, the conversion results in an abstract graph without spatial relationships.
This results in a \textit{natural} graph representation of the dislocation network, hereafter referred to as the discrete dislocation graph as shown in Figure~\ref{fig:R0}.

\subsubsection*{Representation 1 ($R_1$)}\label{{subsubsec:R1}}

\begin{figure}[H]
\centering 
\includegraphics[scale=0.4]{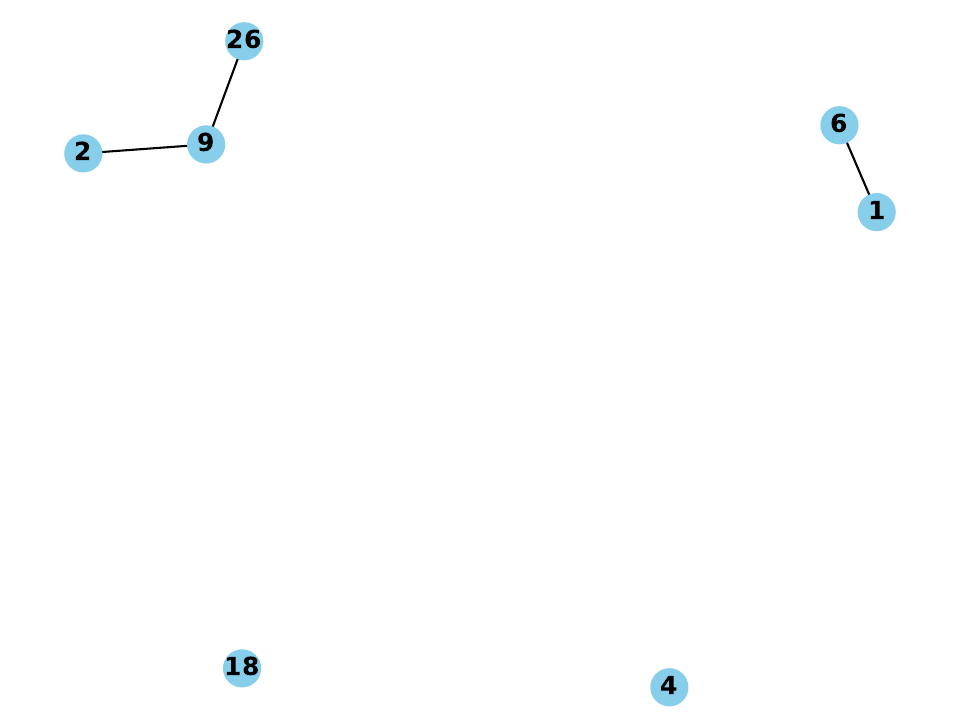}
\caption{In $R_1$, dislocation loops are represented as nodes, while junctions between loops are represented as edges. The three junction edges, $(11,12)$, $(13,14)$ and $(10,9)$, in Figure~\ref{fig:R0} are therefore represented differently: edge $(11,12)$, corresponding to a junction between dislocation loops $2$ and $9$, becomes the edge $(2,9)$; edge $(13,14)$ becomes $(9,26)$; and edge $(10,9)$ becomes $(6,1)$. Thus, each edge in this representation corresponds to a junction in this dislocation network.}
\label{fig:R1}
\end{figure}

An alternative graph representation can be built on the concept of topologically closed dislocation loops~\cite{STRICKER2018319,SUDMANNS2019}.
In this representation, each individual dislocation loop is treated as a single node, while junctions between dislocation loops are represented as edges connecting these nodes.
By abstracting away segment level details that are largely determined by the underlying computational discretizations, this formulation emphasizes the connectivity between dislocation loops and provides a compact description of network topology that conceptually also aligns with CDD~\cite{HOCHRAINER2014167}.
Figure~\ref{fig:R1} shows the dislocation loops as nodes, which are mapped to individual nodes in the corresponding graph representation shown in Figure~\ref{fig:R0}; the junctions (Figure~\ref{fig:R0}) between these loops form the edges connecting the nodes.
While $R_0$ has segment-level resolution, $R_1$ enables connectivity based filtering at the level of entire dislocation loops, allowing specific topological features of the network to be isolated directly.

\subsubsection*{Representation 2 ($R_2$)}\label{subsubsec:R2}
\begin{figure}[H]
\centering 
\includegraphics[scale=0.4]{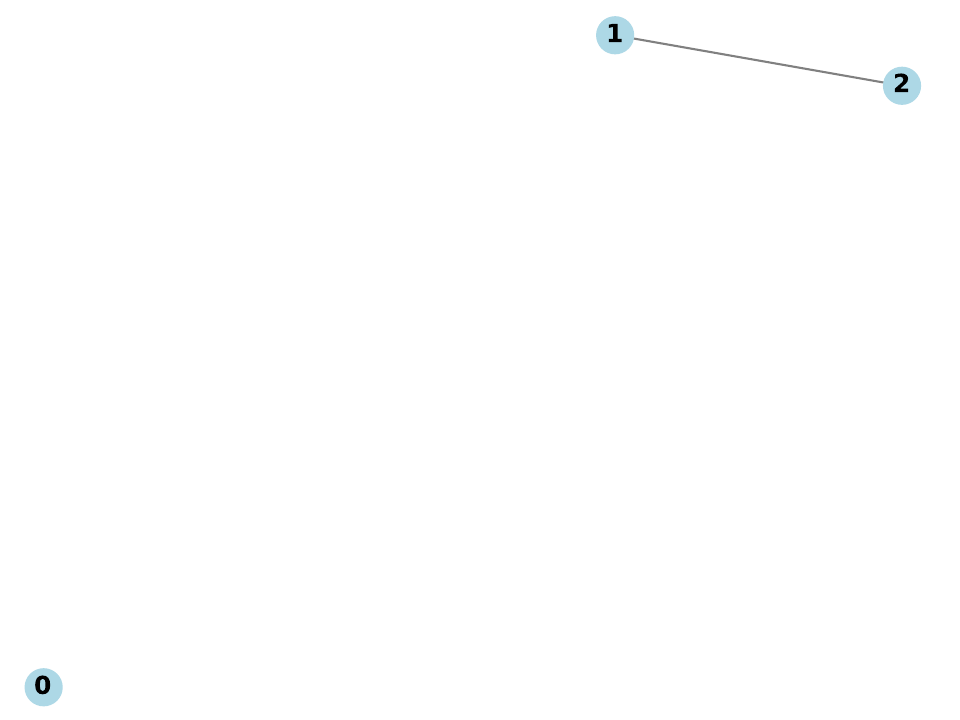}
\caption{In $R_2$, nodes correspond to junctions in the dislocation network, while edges represent the dislocation loops connecting them. Node 1, 2, and 0  correspond to junction edges in Figure~\ref{fig:R0} and~\ref{fig:R1}. The edge $(1,2)$ represents dislocation loop 9 in Figure~\ref{fig:R1}, and node 0 represents the junction between dislocation loops 6 and 1.}
\label{fig:R2}
\end{figure}

A complementary representation to $R_1$ is obtained by reversing the assignment of graph entities: Junctions are interpreted as nodes, dislocations as edges connecting these nodes.
This redefines the network topology as `junction connectivity'.
Figure~\ref{fig:R2} shows the corresponding graph.

\subsubsection*{Representation 3 ($R_3$)}\label{subsubsec:R3}

\begin{figure}[H]
\centering 
\includegraphics[scale=0.4]{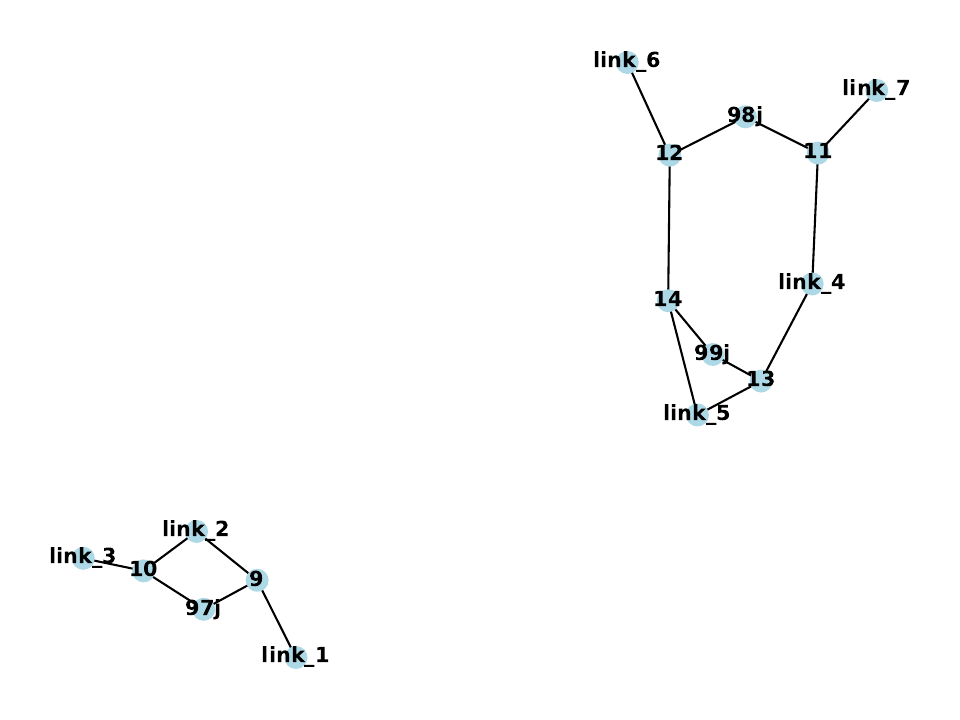}
\caption{$R_3$ corresponds to a hybrid approach, in which higher level nodes include end nodes, link nodes, and junction nodes. The original discrete nodes are treated as lower level entities and are removed through a coarse graining process, with corresponding links introduced to preserve the network connectivity~\cite{KATZER2024115841}. Here, nodes 9, 10, 11, 12, 13 and 14 are end nodes corresponding to the endpoints of junction edges in Figure~\ref{fig:R0}. The junction nodes $97j$, $98j$ and $99j$ represent junctions between dislocation loops, while the link nodes $link_1$-$link_7$ represent extended chains of edges in the dislocation network.}
\label{fig:R3}
\end{figure}

$R_3$ follows the graph formulation introduced by Katzer et al.~\cite{KATZER2024115841}.
In contrast to the preceding representations, their approach distinguishes between three primary node types to retain specific structural features of the dislocation network.
First,~\emph{endnodes} represent the terminal points of dislocation involved in junction formation.
Additional nodes are introduced between these end nodes to explicitly represent the junctions between dislocation loops.
Furthermore, sequences of intermediate nodes along extended chains are combined into single representative node, referred to as \emph{linknodes}.
This abstraction reduces the complexity of the graph while preserving the specific connectivity required to characterize the network and facilitates subsequent data mining analyzes~\cite{KATZER2024115841}.
The resulting graph structure and the different node types are illustrated in Figure~\ref{fig:R3}.

\subsubsection*{Representation 4 ($R_4$)}\label{subsubsec:R4}

$R_4$ is inspired by the junction point density concept introduced by Starkey et al.~\cite{STARKEY2022}. In $R_4$, extended chains of dislocation segments are collapsed into single edges, while the remaining nodes correspond exclusively to junction points.
This abstraction focuses the graph on junction connectivity, motivated by the role of the junctions in constraining dislocation motion and governing the evolution of the dislocation network.
By removing segment level details, $R_4$ reduces the structural complexity of the graph while retaining information associated with the organization and connectivity of junctions.

As shown in Figure~\ref{fig:R0}, all nodes correspond to junctions in the original dislocation network. Open dislocation loops are treated as closed loops, and each extended chain of edges is collapsed in to a single edge.
The original junction edges are preserved in the resulting graph.
Consequently, two distinct types of edges may connected the same pair of nodes, an edge representing a junction in the original network and an edge representing a collapsed chain of segments.
To distinguish these edges, the graph is represented as directed graph, allowing multiple edges between the same pair of nodes.

\begin{figure}[H]
\centering 
\includegraphics[scale=0.4]{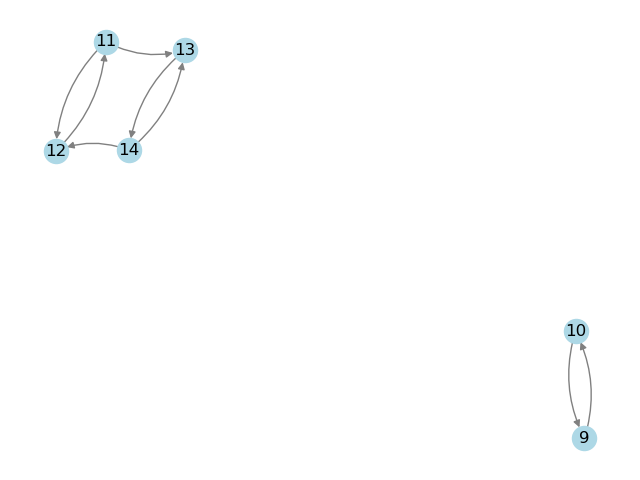}
\caption{In $R_4$, nodes 9, 10, 11, 12, 13, and 14 correspond to the endpoints of junction edge in Figure~\ref{fig:R0}. The original junction edges are preserved, while extended chains of nodes are collapsed into single edges. Collapsed edges are $(11,12)$, $(14,13)$ and $(10,9)$.}
\label{fig:R4}
\end{figure}

\subsection{Graph properties}\label{subsec:Graph_properties}

Once dislocation networks are represented as graphs, a range of graph properties can be evaluated on them.
Python libraries such as NetworkX~\cite{NetworkX} and NetworKit~\cite{staudt2015networkit} provide efficient tools to compute properties at the node, edge, and global levels.
Our focus is on capturing the collective behavior of the dislocation network; therefore, emphasis is placed on global graph properties.
We evaluate the average degree, transitivity, assortativity, graph density, diameter, and the total number of nodes and edges. Detailed definitions and interpretations of these properties are provided in the following.

\subsubsection*{Average degree distribution}\label{subsubsec:avg_DD}

The degree distribution~\cite{BarabasiNetworkScience, Newman2003} of a node is defined as the number of edges connected to it, representing the local connectivity within the graph. Accordingly, the degree distribution characterizes how these connections are distributed across all nodes in the graph.
While degree is inherently a node level property, its distribution provides insight into the structural organization of the graph.
To obtain a global graph property of connectivity, we compute the average degree distribution is over all nodes which provides a measure for the overall level of interconnectedness.

\subsubsection*{Transitivity}\label{subsubsec:transitivity}

Transitivity provides a measure of the overall triangle density in a graph and is defined as the ratio of the total number of triangles to the number of connected triples of nodes, as given in Eq.~\ref{eq:Transitivity}.
It is a global property of the graph and is often referred to as the global clustering coefficient~\cite{BarabasiNetworkScience}.

A key distinction between transitivity and the average clustering coefficient lies in how node contributions are weighted.
In the average clustering coefficient, each node contributes equally, making the measure sensitive to low degree nodes and allows to identify `local community structure'.
In contrast, transitivity weights nodes according to their participation in connected triples, which gives more weight to high degree nodes.
As a result, transitivity is dominated by highly connected regions of the network and provides a measure of the overall `global cohesion'~\cite{Newman2003, newman2010}.

\begin{equation}
\mathrm{Transitivity} = \frac{3\cdot \text{ number of triangles in the network}}{\text{number of connected triples of nodes}} 
\label{eq:Transitivity}
\end{equation}
  
\subsubsection*{Assortativity}\label{subsubsec:assortativity}

Assortativity is a property that quantifies the tendency of nodes to connect with other nodes that are similar with respect to a given attribute, most commonly node degree which can be calculated by Eq.~\ref{eq:Assortativity}~\cite{Newman2002Assortativity}.
Degree assortativity, in particular, captures whether high degree nodes preferentially connect to other high degree nodes, and similarly for low degree nodes.
Assortativity provides a measure for the structural organization and homogeneity of the network.

A positive assortativity coefficient indicates assortative mixing, where nodes tend to connect to others with similar properties, reflecting a more homogeneous graph.
Negative values correspond to disassortative mixing, where nodes preferentially connect to dissimilar nodes, indicating a more heterogeneous network topology.
This distinction is a measure for the organization of connectivity how interactions are distributed across the graph~\cite{Newman2002Assortativity}.

\begin{equation}
\mathrm{r}=\frac{\sum_{jk}{(e_{jk\ \ }-\ q_jq_k)}}{\sigma_q^2},
\label{eq:Assortativity}
\end{equation}

 where $e_{jk}$ is the fraction of edges connecting to nodes of degree $j$ and $k$, $q_k$ is the probability that a randomly chosen edge is attached to a node of degree $k$, and $\sigma_q^2$ is the variance of the degree distribution over edges.
 
\subsubsection*{Graph density}\label{subsubsec:graph_density}

Graph density is defined as the ratio of the number of edges present in the network to the maximum possible number of edges between nodes and provides a normalized measure of how densely the network is connected, independent of its size.
For an undirected graph, the maximum number of possible edges is determined by the total number of node pairs.

This metric reflects the overall level of connectivity in the network, where higher values indicate a more densely interconnected graph, while lower values correspond to sparser graphs~\cite{newman2010}. The graph density is calculated differently for undirected and directed graphs, as the maximum number of possible edges differs between the two graph types. The corresponding graph density for undirected and directed graph are defined in Eq.~\ref{eq:undirected_graph_density} and Eq.~\ref{eq:directed_graph_density}, respectively, as follows:

\begin{figure}[ht]
\centering
\begin{minipage}{0.48\textwidth}
\begin{equation}
\mathrm{Density} = \frac{2E}{N \cdot (N - 1)}
\label{eq:undirected_graph_density}
\end{equation}
\centering
\end{minipage}
\hfill
\begin{minipage}{0.48\textwidth}
\begin{equation}
\mathrm{Density} = \frac{E}{N \cdot (N - 1)}
\label{eq:directed_graph_density}
\end{equation}
\centering
\end{minipage}
\end{figure}

where $E$ is total number of edges in graph and $N$ is total number of nodes in graph.

\subsubsection*{Diameter}\label{subsubsec:diameter}

The diameter~\cite{BarabasiNetworkScience} of a graph is defined as the maximum distance between any pair of nodes, representing the ``longest shortest path'' in a graph.
In other words, it is the largest value among the shortest path lengths between all pairs of nodes~\cite{staudt2015networkit,CRESCENZI201384}.
As a global property, the diameter provides measures the overall extent of the network and its connectivity structure.
A larger diameter indicates a more extended or weakly connected network, whereas a smaller diameter reflects a more compact structure in which nodes are more closely linked~\cite{newman2010}.

\subsubsection*{Number of nodes}\label{subsubsec:num_nodes}

The number of nodes is a fundamental measure of a graph, its size.

\subsubsection*{Number of edges}\label{subsubsec:num_edges}

Similarly fundamental, the number of edges denotes the total number of connections in a graph.

\subsection{Embedding}\label{subsec:embeddding}

\begin{figure}[H]
\begin{center}
\begin{tikzpicture}

    \node at (-7,0) {\includegraphics[width=5.5cm]{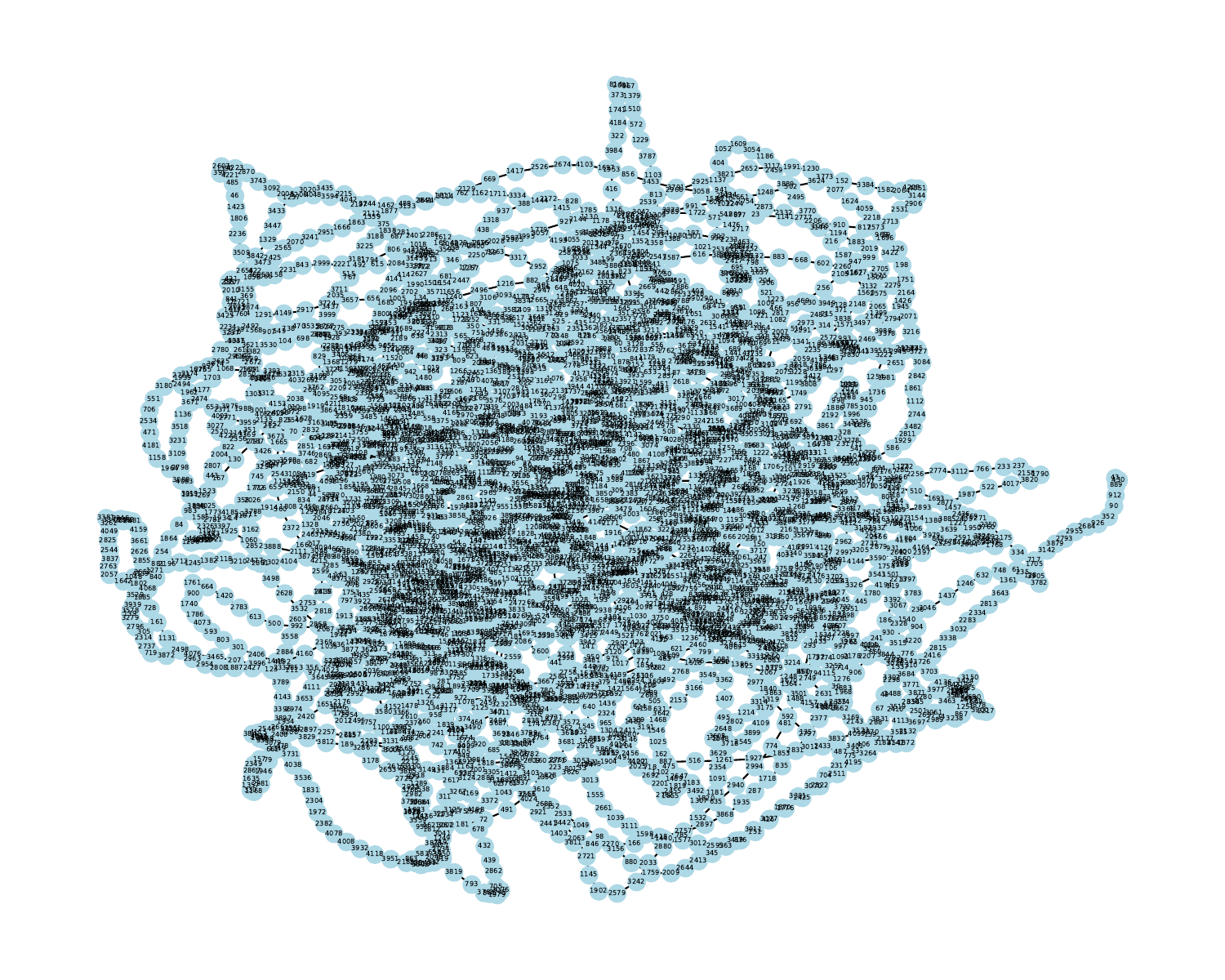}}; 

    \draw[->, thick] (-4.5,0) -- (-1.8,0) node[midway, above] {Graph properties};

    \node[draw] (vector) at (0.8,0) {\(\begin{bmatrix}
        \text{Strain rate} \\
        \text{Average degree} \\
        \text{Transitivity} \\
        \text{Assortativity} \\
        \text{Graph density} \\
        \text{Diameter} \\
        \text{Number of nodes} \\
        \text{Number of edges}
    \end{bmatrix}\)};

\end{tikzpicture}
\end{center}
\caption{Schematic of the process for evaluating graph properties of a spatially resolved dislocation network in form of an abstract graph (left). The data, initially high dimensional, is transformed into a lower dimensional vector representation (right).}
\label{fig:schemcatic_vector}
\end{figure}

For each snapshot along every discrete dislocation trajectory, we evaluate all graph properties in all representations $R_i$.
This evaluation yields a global graph property vector for each snapshot and representation $R_i$ that characterizes the corresponding dislocation network.
Together with the associated mechanical properties, we use these 8-dimensional vectors to characterize the evolution of the dislocation structures.
To explore the relationships among these networks in their different representations, we employ dimensionality reduction technique by converting the high-dimensional vector representation (strain rate, average degree, transitivity, assortativity, graph density, diameter,  number of nodes, and number of edges) into lower dimensional embeddings (``maps'').
Figure~\ref{fig:schemcatic_vector} illustrates this transformation from an abstract graph to its vector representation.
The lower-dimensional embeddings are useful to assess the relationships of different dislocation networks and we use them to map to conventional dislocation structure characteristics like associated stress and dislocation density.

Each row of the data set, which get constructed by converting the dislocation networks into graph using the different graph representations (subsection~\ref{subsec:graph_R}) and calculating the the graph properties (subsection~\ref{subsec:Graph_properties}), represents the mechanical and graph properties associated with a snapshot of the dislocation network recorded periodically throughout DDD simulations, resulting in a high dimensional description of the evolution of the network. The primary objective of embedding is to encode these observations' relationships into lower dimensional vectors enabling easier visualization and analysis of complex datasets.
Methods like principal components analysis (PCA)~\cite{Pedregosa_Scikit-learn_Machine_Learning_2011} help highlight interactions between features like stress and dislocation density, and the identification of trends, outliers, and groups within the representation.

\subsection{Random forest regression model}\label{random_forest_model}

To assess the extent to which the graph representations capture information relevant to the mechanical response, we employ a standard random forest regression model~\cite{Breimam_2001}, as implemented in scikit-learn~\cite{Pedregosa_Scikit-learn_Machine_Learning_2011}. We use a simple random forest regression as our predictive model for three reasons.
First, the functional relationship between network topology and mechanical responses is not known a prior; indeed, it is not yet established whether the selected graph properties contain sufficient information to predict the mechanical response.
As a non-parametric method, random forest regression does not require an assumed functional form.
Second, the method is robust and stable, bootstrap aggregation over many tress controls variance and makes the model resistant to overfitting, so it performs reliably out of the box on a dataset of this size without extensive hyper parameter tuning.
Third, is it a well established, widely use ensemble methods whose behavior and evaluation are well understood, and which additionally yields feature importance information that allows to asses whether dislocation network topology, combined with the strain rate, predicts yield stress, a foundation from which richer descriptors, the full loading history, and further modeling can be pursued in subsequent work.

For each dislocation network, the model predicts the logarithm of the yield stress using the corresponding graph property vector together with the imposed strain rate as input features. The random forest consists of 500 regression trees, each trained of a bootstrap resample of the training data, and the final prediction is obtained by averaging the predictions across all trees.
Apart from the number of estimators, all hyper parameters are kept at their scikit-learn default value, and no additional hyperparameter tuning is performed. 
Model performance is assessed by strain rate stratified five-fold cross-validation.
The runs are partitioned into five groups stratified by strain rate, so that each fold, and therefore each training set, contains runs from the full range of loading rates.
Predictions are generated for out of fold data points (each run predicts stress that did not see it), and the coefficient of determination $R^2$ is computed over all out-of-fold predictions.
Rows corresponding to graph representations with zero value for the considered graph properties were excluded from the training dataset.

\section{Results}\label{sec:results}

A total of 627 discrete dislocation dynamics (DDD) simulations were performed to systematically investigate the effects of size, crystallographic orientation, initial dislocation density, and applied strain rate on the yielding response.
The simulation data set spans four sizes (side length of square cross section), namely 0.5$\upmu$m, 1$\upmu$m, 1.5$\upmu$m and 2$\upmu$m, all with a fixed aspect ratio of 1:3.
For each size, uniaxial tension is applied along four crystallographic orientations [100], [110], [111], and [234] in an FCC crystal.
For each orientation, a range of initial dislocation densities spanning approximately $10^{11}~$m$^{-2}$ to $10^{14}~$m$^{-2}$ is used.
The initial density is created using Frank-Read sources equally distributed over all slip systems with an average length corresponding to the mean dislocation distance $\bar{d} = 1/\sqrt{\rho_{\text{start}}}$ with a small random variation.
In addition, seven applied strain rates are investigated, covering the range from $5.0~$s$^{-1}$ to $5.0\times 10^{6}~$s$^{-1}$.
This systematic variation in simulation parameters results in a comprehensive data set covering a broad range of microstructural and loading conditions. 

To assess the applicability of the universal scaling relation proposed by Fan et al.~\cite{Fan2021} w.r.t. our DDD simulations, the axial yield stress $\sigma_y$ is extracted from the each simulation.
For the definition of the yield stress, we follow~\cite{Fan2021}: By default, the yield stress is defined at a plastic strain of $\varepsilon_y^p$ = 0.2 \%.
For the two lowest strain rates, $5.0$~$s^{-1}$ and 50~$s^{-1}$, a lower plastic strain offset of 0.05~\% is used.

\begin{figure}[H]
    \centering
    \begin{minipage}{0.49\textwidth}
        \centering
        \includegraphics[width=\linewidth]{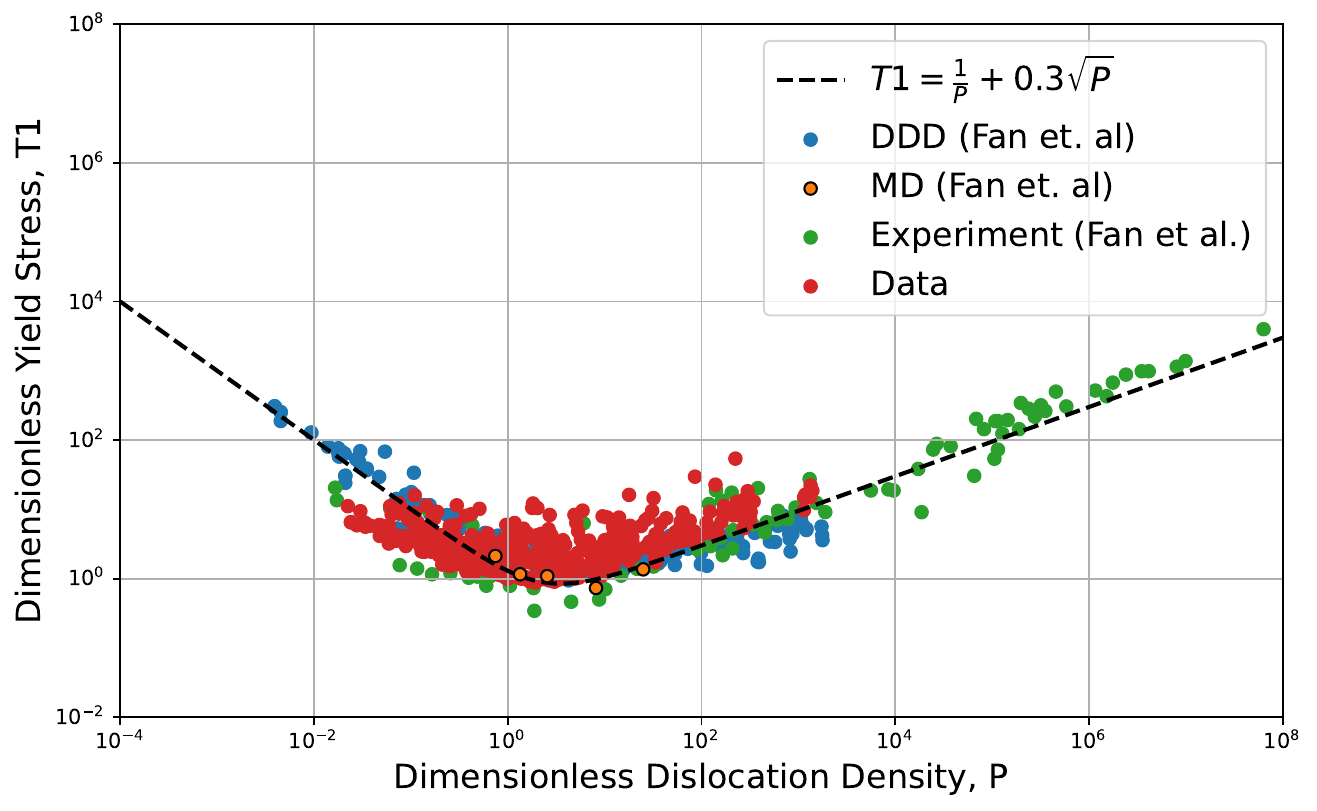}
    \end{minipage}
    \hfill
    \begin{minipage}{0.49\textwidth}
        \centering
        \includegraphics[width=\linewidth]{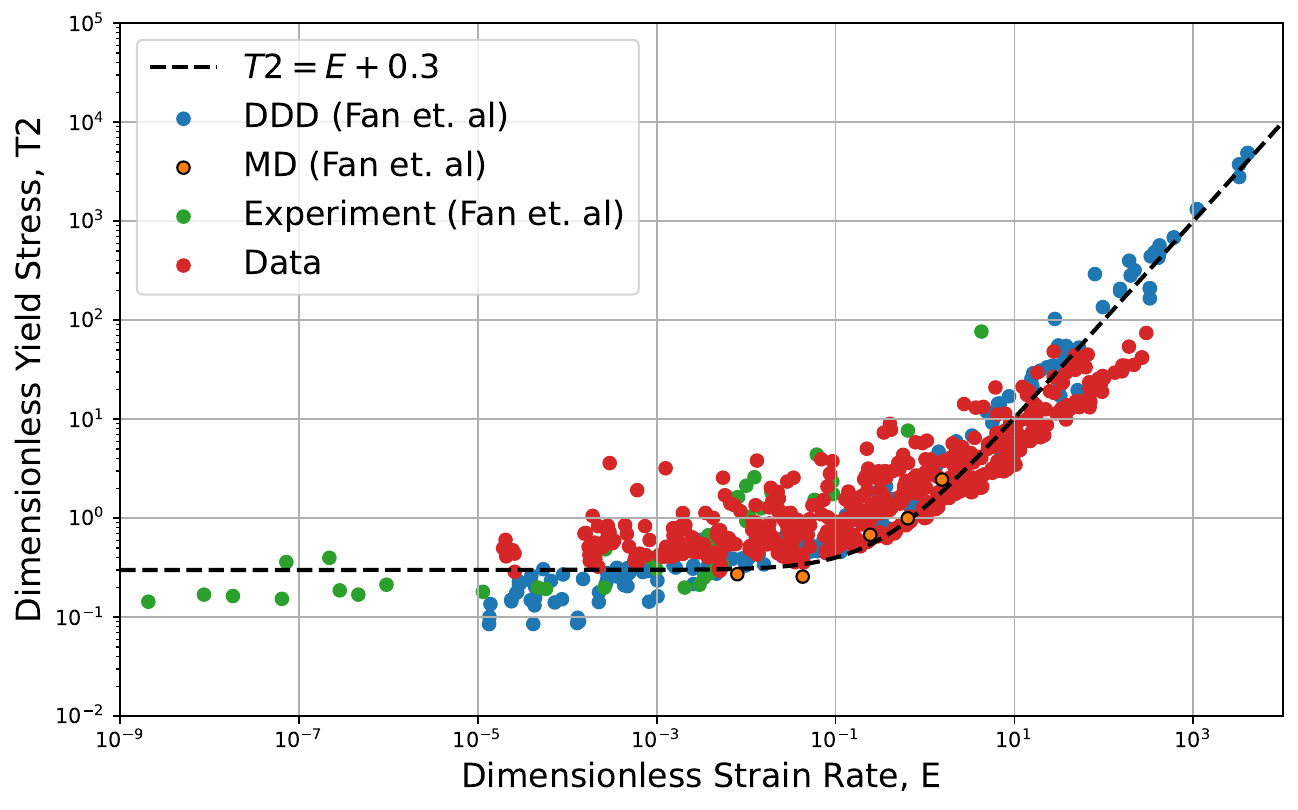}
    \end{minipage}
    
    \caption{Dimensionless yield stress as a function of dimensionless dislocation density (left) and dimensionless strain rate (right), after~\cite{Fan2021}; our data is shown in red. Data from~\cite{Fan2021} are group according to their origin as reported there.}
    \label{fig:fan_vs_project_data}
\end{figure}

The resulting yield point data are subsequently compared with the universal scaling curve proposed from~\cite{Fan2021} (cf. Figure~\ref{fig:fan_vs_project_data}).
The yield stress obtained from our DDD simulations (red data points) show very good agreement with and are distributed closely around the data of~\cite{Fan2021}.
Each red data point represents the yield response obtained from an individual DDD simulation, thereby corresponding to a distinct dislocation network and its associated combination of size, crystallographic orientations, initial dislocation density and applied strain rate.
Close correspondence between our simulation data and the `universal curve'~\cite{Fan2021} suggests that the proposed scaling relationship captures the dominant trends in yielding behavior across the range of microstrucral and loading conditions investigated here.
 
\begin{figure}[H]
    \centering 
    \includegraphics[
        width=\textwidth,
        keepaspectratio
    ]{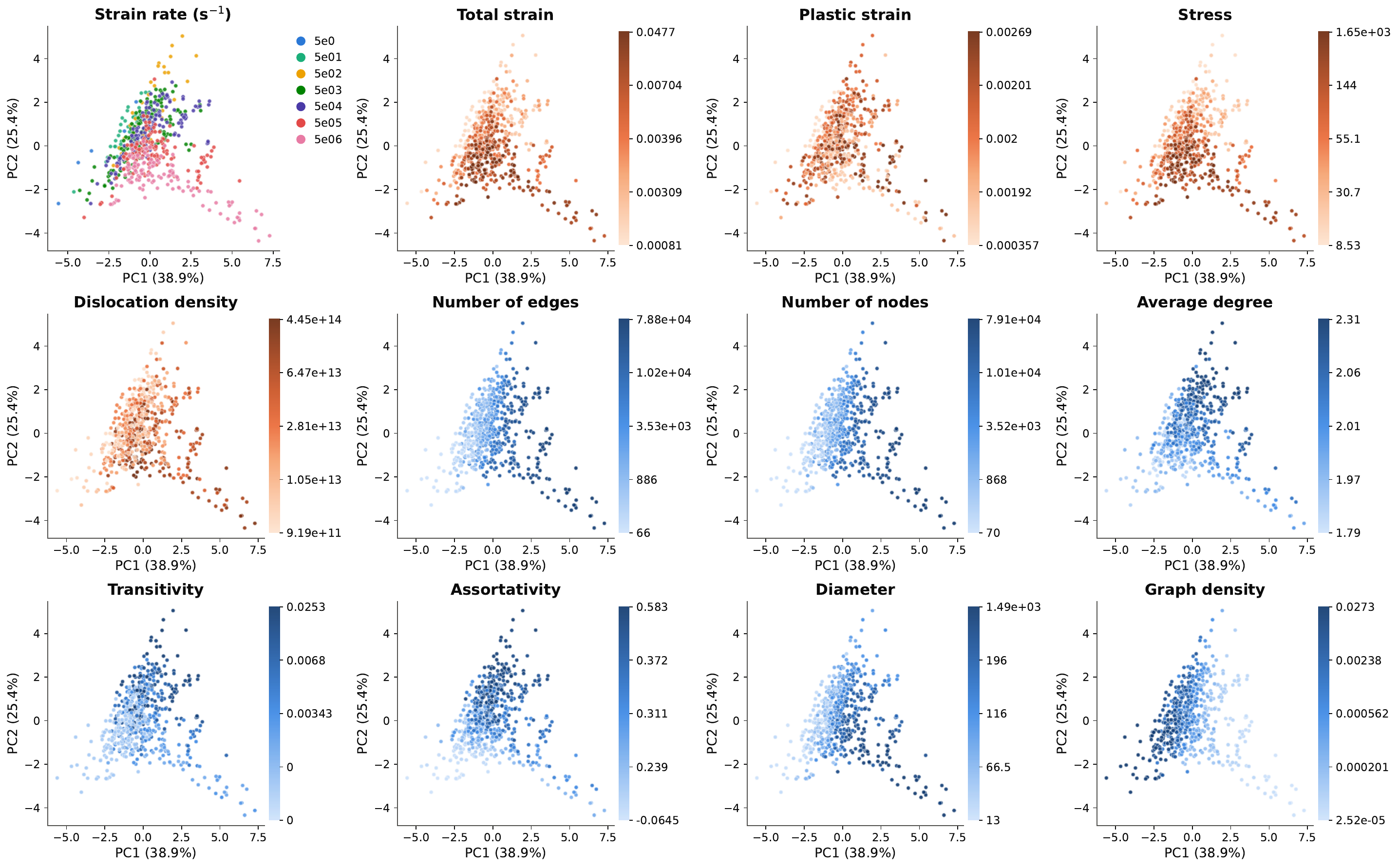}
    \caption{Two-dimensional principal component analysis embedding of the dislocation network in the $R_0$ graph representation. The same embedding is show colored by each mechanical property (yield stress, total strain, plastic strain, dislocation density) and each graph property (number of nodes, number of edges, average degree, transitivity, assortativity, diameter, graph density), as well as the applied strain rate.}
    \label{fig:pca_R0}
\end{figure}

\begin{figure}[H]
    \centering 
    \includegraphics[
        width=\textwidth,
        keepaspectratio
    ]{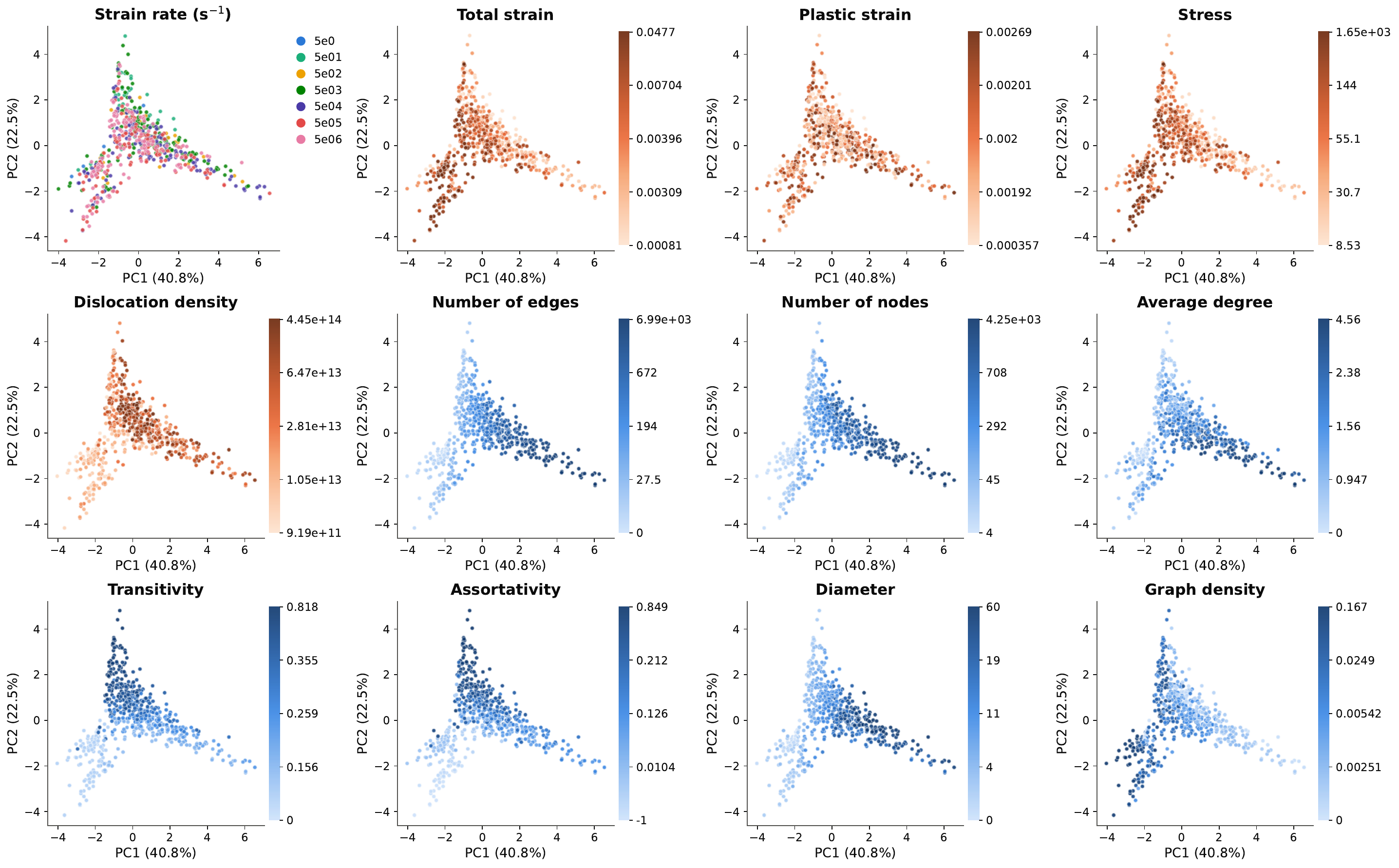}
    \caption{Two-dimensional PCA embedding of the dislocation networks in the $R_1$ graph representation, with the same embedding Colored by each mechanical property (yield stress, total strain, plastic strain, dislocation density), each graph property (number of nodes, number of edges, average degree, transitivity, assortativity, diameter, graph density), and the applied strain rate.}
    \label{fig:pca_R1}
\end{figure}

\begin{figure}[H]
    \centering 
    \includegraphics[
        width=\textwidth,
        keepaspectratio]{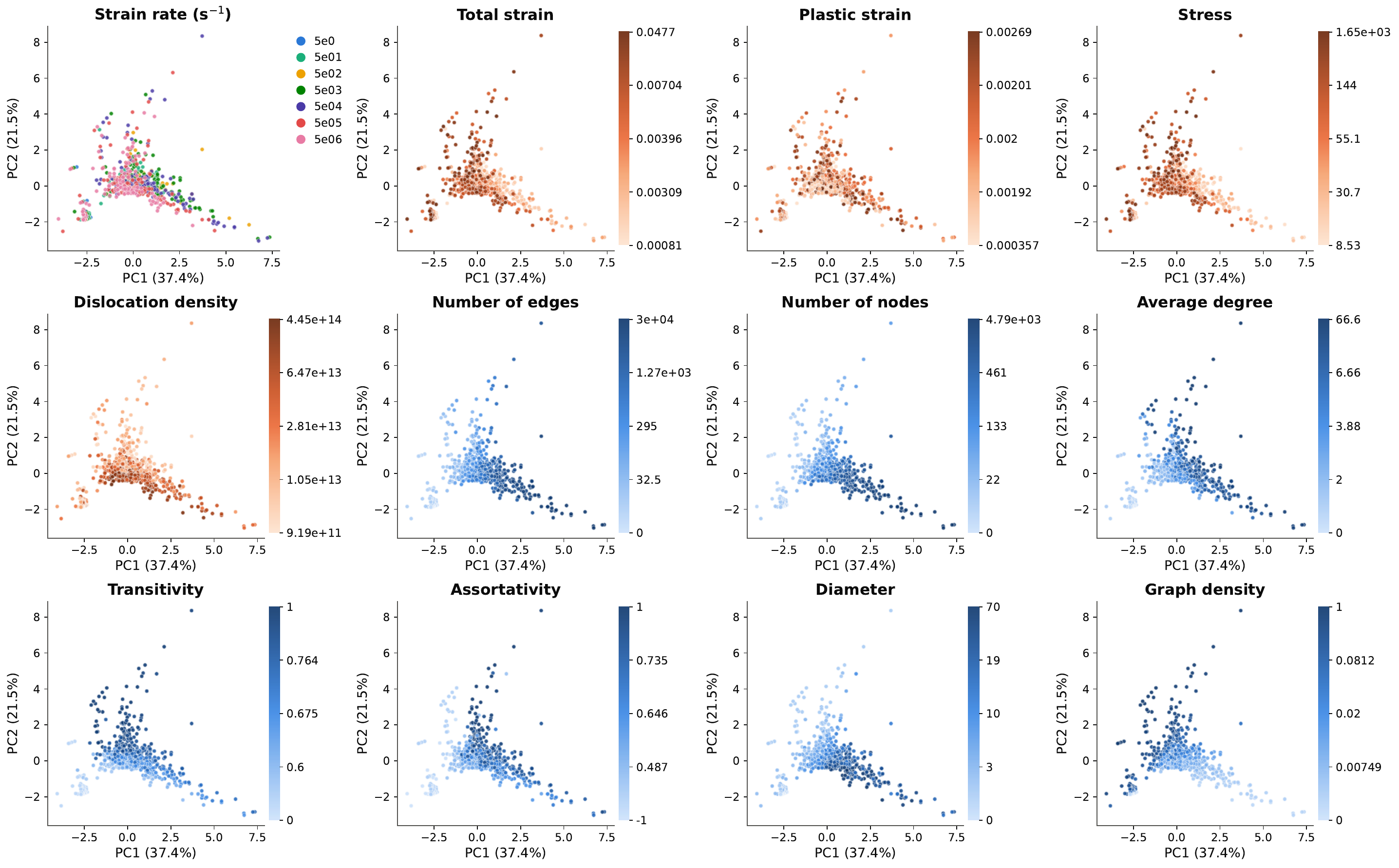}
    \caption{Two-dimensional PCA embedding of the dislocation network in the $R_2$ graph representation colored by each mechanical property, each graph property, and applied strain rate.}
    \label{fig:pca_R2}
\end{figure}

\begin{figure}[H]
    \centering 
    \includegraphics[
        width=\textwidth,
        keepaspectratio
    ]{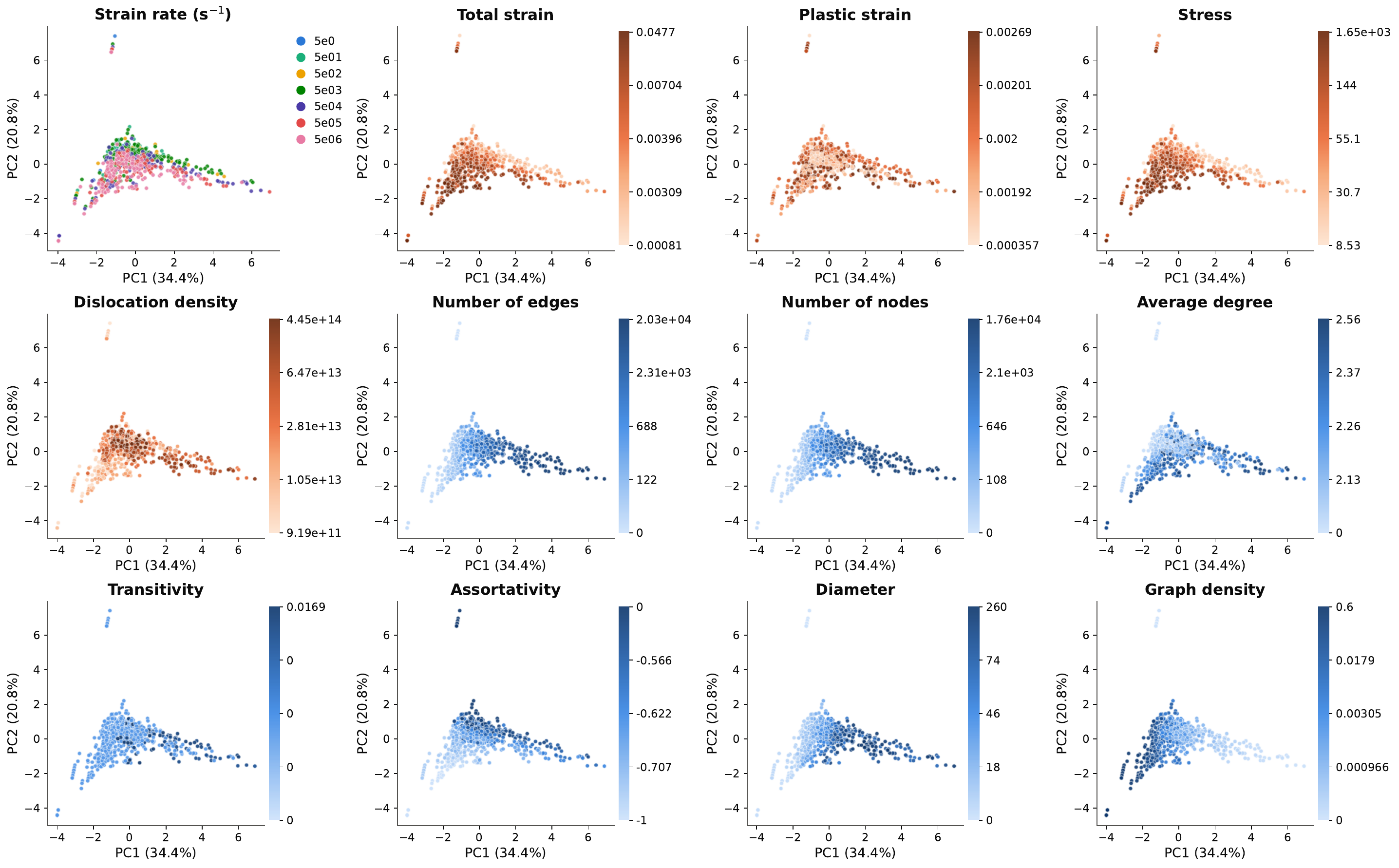}
    \caption{Two-dimensional PCA embedding of the dislocation networks in the $R_3$ graph representation, with the same embedding colored by each mechanical property (yield stress, total strain plastic strain, dislocation density), each graph property (number of nodes, number of edges, average degree, transitivity, assortativity, diameter, graph density), and the applied strain rate.}
    \label{fig:pca_R3}
\end{figure}

\begin{figure}[H]
    \centering 
    \includegraphics[
        width=\textwidth,
        keepaspectratio
    ]{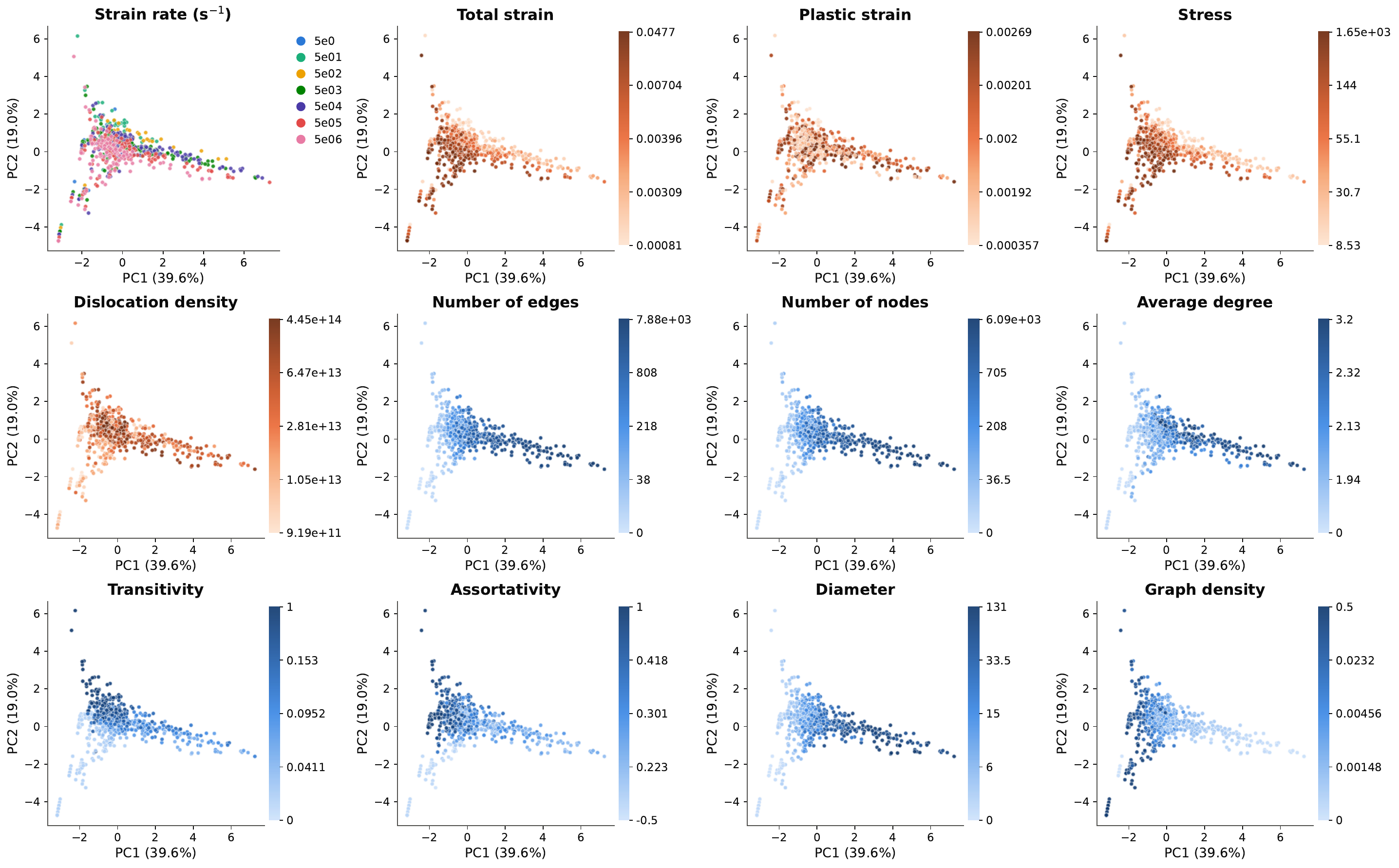}
     \caption{Two-dimensional PCA embedding of the dislocation network in the $R_4$ graph representation, show colored by each mechanical property (yield stress, total strain, plastic strain, dislocation density), each graph property (number of nodes, number of edges, average degree, transitivity, assortativity, diameter, graph density), and the applied strain rate.}
    \label{fig:pca_R4}
\end{figure}

Following the conventional analysis, each dislocation network in our dataset is converted into a graphs using the five representations $R_i$ introduced in Section~\ref{subsec:graph_R}.
For every resulting graph, the topological properties defined in Section~\ref{subsec:Graph_properties} are evaluated.
The feature matrix used in the following analysis contains only the topological properties together with the applied strain rate; no mechanical properties (stress, strain) are included as inputs.
All graph properties are standardized to zero means and unit variance prior to dimensionality reduction, so that properties of differing magnitude contribute comparably.
Principal component analysis (PCA) is then applied separately to each representation $R_i$, and the resulting two dimensional embeddings are shown in Figure~\ref{fig:pca_R0} to Figure~\ref{fig:pca_R4}.
In each figure the respective embedding coordinates (principal components 1 and 2) are reproduced several times but colored by a different graph properties and some of the mechanical properties.
This allows to asses how each graph property varies with the two principal component axes.

Across all the five representations the embeddings show clear trends: the configurations arrange themselves along well defined directions in the reduced space, indicating that the topological properties are strongly correlated with one another and that the intrinsic dimensionality~\cite{Facco_2017} of the graph properties set is considerably lower than its nominal dimensionality. Indeed, the TwoNN estimator~\cite{Facco_2017} gives an intrinsic dimension of only $\approx 3$ for seven graph properties, a value that stay stable when the estimate is repeated on random subsets of the data.

The embedding of $R_0$ (Figure~\ref{fig:pca_R0}) shows the clearest separation of graph properties groups among the five representations: The number of nodes, number of edges, and diameter vary along a common direction closely aligned with the first principal component, and therefore jointly describe the overall size and spatial extent of the network.
The average degree, transitivity, and assortativity form a second tightly aligned group with a gradient oblique to the first, indicating that local connectivity, clustering, and degree correlation vary together and largely independently of network size.
Graph density varies in the opposite direction to the size-related graph property group, as required by its definition.
Strain rate is also well resolved in this embedding space: configurations at different imposed rates occupy distinguishable regions of the plane, with the highest rates displaced towards large PC1 and low PC2.

Among the overlaid mechanical quantities, stress and total strain both exhibit well defined gradients, oriented approximately along PC2 and therefore oblique to both graph property groups.
The dislocation density shows a weaker but still discernible gradient. In contrast, the plastic strain does not show a clear trend in embedding space.

For $R_1$ (Figure~\ref{fig:pca_R1}), the point cloud is no longer approximately convex but instead develops distinct branches, accompanied by a corresponding change in the grouping of the properties. The number of nodes, number of edges, diameter, and, in this representation, also the average degree vary along a common direction. Thus, the average degree in $R_1$ is governed primarily by the overall size of the network rather than by its local connectivity. This is the main difference from $R_0$, where the average degree is associated with the cluster-related quantities.
Transitivity and assortativity form a separate and mutually consistent pair, with a gradient approximately along PC2, while the graph density varies in the opposite direction to the size-related group.

The lower left branch of the embedding space contains configurations with very small node and edge counts, for which the average degree, transitivity, and diameter approach zero, while assortativity approaches $-1$.
This branch represents a collection of weakly connected dislocation networks, rather than a proper network and should be interpreted with caution.
The corresponding states are also associated with elevated stress and total strain.
Consequently, the apparent correlation between the mechanical state and the topological properties in $R_1$ is at least partly influenced by network fragmentation at large deformations and is quantitatively weaker than the corresponding gradients in $R_0$.

The embedding of $R_2$ (Figure~\ref{fig:pca_R2}) reproduces the overall embedding structure and correlations observed in the preceding cases.
The number of nodes, number of edges, and diameter again form a size related group, while the graph density varies in the opposite direction.
The average degree, transitivity, and assortativity are again associated, recovering the qualitative behavior of $R_0$ rather than $R_1$.
Their alignment, however, is noticeably less pronounced than in $R_0$, with the transitivity appreciably rotated relative to the other two.
Graph density is approximately orthogonal to this connectivity related group and antiparallel to the size related group, consistent with the other representations.

As in $R_1$, the graph property ranges reveal near-degenerate configurations, i.e. graphs with vanishing node and edge counts at one extreme and complete subgraphs at the other.
Both transitivity and graph density are unity in those cases.
The mechanical property overlays are only weakly correlated in this embedding: stress, total strain, and dislocation density all exhibit faint gradients of comparable magnitude, and none of them are resolved as clearly as in $R_0$ or $R_1$.
$R_2$ is therefore the representation in which the topological properties provide the least information about the mechanical state.

For $R_3$ (Figure~\ref{fig:pca_R3}), the size-related properties again behave consistently: the number of nodes, number of edges, and diameter share a common gradient, while the graph density varies in the opposite direction.
The distinguishing feature of this representation lies in the remaining graph properties.
The average degree and assortativity vary along markedly different directions, i.e. in $R_3$ they act as independent graph properties rather than collapsing onto a common gradient as in $R_0$ and $R_2$.
In contrast, the transitivity is essentially constant and remains close to zero for nearly all configurations, so its coloring shows no resolvable variation.
This is a property of the representation itself, which produces graphs that are almost free of closed triangles, and therefore the transitivity does not provide any additional information in $R_3$.

Consistent with this reduced redundancy among the properties, $R_3$ has the lowest fraction of variance captured by the first two components among all five representations, indicating a higher intrinsic dimensionality of its property space.
Stress and total strain both show clear trends with gradients that agree to within a few degrees.
This agreement is expected, since stress typically increases monotonically with accumulated strain over the deformation range considered, and therefore does not by itself demonstrate that the graph properties encode the mechanical response.
Rather, the relevant observation is that the common stress-strain direction is well defined in $R_3$, whereas it is diffuse in $R_2$.

Finally, for $R_4$ (Figure~\ref{fig:pca_R4}), the number of nodes, number of edges, and diameter all vary along a common direction, making the grouping of properties closely resemble that of $R_1$.
Transitivity, assortativity, and average degree are again associated with one another, but their gradients are not collinear.
Their directions are broadly similar, but differ noticeably in orientation, indicating that clustering and degree correlation are only partially coupled in this representation.
Graph density once again varies antiparallel to the size-related group.
Stress and total strain exhibit visible gradients along a common direction, comparable to those observed for $R_3$, although somewhat less clearly defined.

\begin{figure}[H]
    \centering 
    \includegraphics[
        width=\textwidth,
        keepaspectratio
    ]{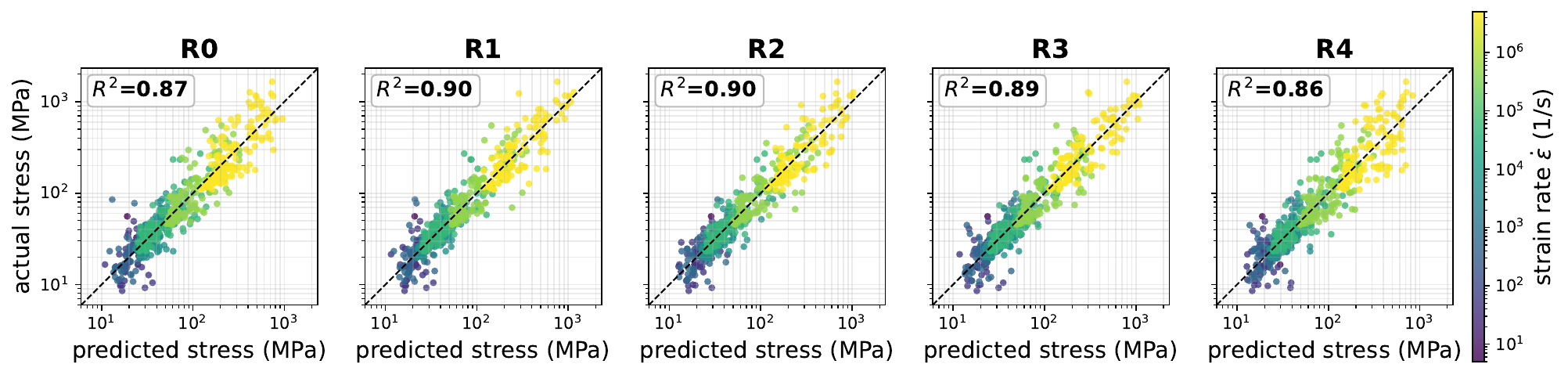}
    \caption{True vs. predicted yield stress for each of the five graph representation ($R_0$-$R_4$), using a standard random forest model trained on the five best graph properties of the dislocation network together with the strain rate. Predictions are out-of-fold values from the strain rate stratified five-fold cross-validation over the full dataset (all sizes). Each data point is one configuration, colored by its applied strain rate (color bar); the dashed line (angle bisector) marks perfect prediction, and the per-panel $R^2$ is the out-of-fold coefficient of determination. Both axes are logarithmic. Accuracy is high and consistent across representations ($R^2\approx 0.86-0.90$).}
    \label{fig:random_forest_all_sizes}
\end{figure}

The embedding analyses indicate that representing the dislocation network as an abstract graph and computing its graph properties capture information relevant to some mechanical properties.
We now test this directly using these graph properties to predict the yield stress.
For each of the five graph presentations ($R_0$-$R_4$), a standard random forest model is trained on the network properties together with the imposed strain rate and evaluated by the strain rate stratified five-fold cross-validation.
Figure~\ref{fig:random_forest_all_sizes} shows the out-of-fold predictions of yield stress against the true yield stress for all sizes.
All five representations follow the angle bisector closely, with cross validation $R^2$ between 0.86 and 0.90, confirming that the graph properties of the dislocation network, combined with the strain rate, predict the yield stress with good accuracy. For comparison, a baseline using only the strain rate and the dislocation density reaches $R^2 \approx 0.86$, so the graph properties improve on it for $R_0$-$R_3$ ($R^2 \approx 0.87-0.90$) but not for $R_4$.

\begin{figure}[H]
    \centering
    \includegraphics[
        width=\textwidth,
        keepaspectratio
    ]{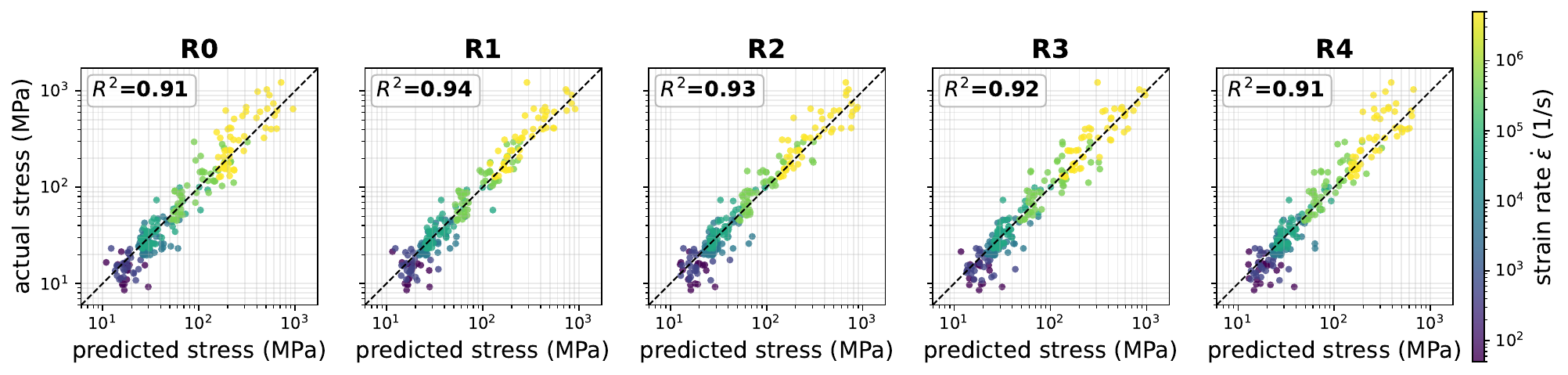}
    \caption{Predication accuracy for the larger networks (size 3 (1.5$\upmu$m) and size 4 (2$\upmu$m)), shown separately for each representations ($R_0$-$R_4$). A single random forest is trained on whole dataset (all size) with the five best graph properties plus strain rate, under, strain rate stratified five-fold cross-validation; out-of-fold predictions are then restricted to the size 3 and size 4 configurations. Each point is one such configuration, colored by its applied strain rate (colored bar); the dashed line is the 1:1 reference and the per panel $R^2$ is evaluated on the size 3$\&$4 subset. Both axes are logarithmic. Restricting the shared all sizes model to the larger size gives $R^2\approx0.91\text{-}0.94$, slightly above its all sizes accuracy larger networks are predicted more reliably.}
    \label{fig:random_forest_3&4_size}
\end{figure}

Restricting the analysis to the larger simulation boxes (sizes 1.5$\upmu$m and 2$\upmu$m) further improves the prediction (Figure~\ref{fig:random_forest_3&4_size}), the predicted and true stress collapse even more closely onto the angle bisector, and cross-validation $R^2$ increases to $0.91$-$0.94$.
This stronger performance on larger networks can be rationalized: Larger simulations typically have more dislocations and therefore a larger network, i.e. network measures \textit{make more sense} for an actual network.

\section{Discussion}\label{sec:discussion}

From some representations the visible trend of certain mechanical properties align with those of a graph properties; $R_1$ in particular shows a clear correlation between the graph properties and the mechanical quantities.
Three features of the embedding are consistent across the analysis.

First, the separation of the graph properties space into a size-dominated and connectivity-dominated direction is reproduced across all representations, but the assignment of individual properties to these direction depends on the representation.
The first principal component is governed by the size/spatial extent properties (number of node, number of edge and diameter), with graph density opposing this group, while the second component collects the local connectivity and ordering properties (average degree, transitivity, and assortativity).
The relationship of the average degree with the other graph properties varies across the different representations. In $R_0$ and $R_2$, the average degree is associated more closely with clustering related properties, whereas in $R_1$ and $R_4$ it shows greater similarity to network size related properties. In $R_3$, the average degree appears comparatively independent of other graph properties.

Second, the total strain and stress shows a clear trend in all five embedding spaces, whereas plastic strain, which is mutually redundant over the range considered (r$~\approx 0.96$), is resolved to different degrees.
The cleanest alignment between the mechanical and topological properties is observed for $R_1$ (Figure~\ref{fig:pca_R1}) and $R_3$ (Figure~\ref{fig:pca_R3}), whereas $R_0$ (Figure~\ref{fig:pca_R0}) shows essentially no alignment between the two.

Third, $R_3$ combines the lowest two component variance fraction with the largest number of mutually independent property directions; transitivity, in particular, is uninformative in the leading plane of this representation, contributing almost nothing to the first two components and instead defining a separate axis of its own.

Taken together, although seven topological graph properties are computed for every network, they do not vary independently.
When the strain rate is added, the first two principal components account for between 54.6\% and 64.3\% of the property variance (62.9\%-71.5\% for the graph properties alone), and the seven graph properties consistently fall into three groups, a size group, a local connectivity group and graph density, which largely opposes one of the other two. 
These three group suggests that a dislocation network is describe mainly by how large it is and how its segments are locally connected, with graph density following from these two rather than adding independent information.
Because the membership of these groups slightly changes between representations, the choice of representation decide which properties carry the connectivity information, which implies that a compact graph property set, such as one property per group, may be sufficient but should be chosen for each representation separately.
This indicates that not all seven graph properties are required and supports the use of a standard random forest model evaluated with five-fold cross-validation.

Across all five graph representations, a standard random forest using each representation's most suitable five graph properties together with the strain rate, a total of 6 dimensions, predicts the yield stress with high accuracy under strain rate stratified five-fold cross-validation.
On the full dataset (all box sizes) the out-of-fold coefficient of determination lies between $R^2$ = 0.86 and 0.9 ($R_0$ 0.87, $R_1$ 0.90, $R_2$ 0.90, $R_3$ 0.89, $R_4$ 0.86) with $R_1$ the strongest representations.
When the analysis is restricted to the larger simulations (size 1.5$\upmu$m and 2$\upmu$m) the accuracy rises consistently to $R^2 = 0.91-0.94$ across every representation.
The larger networks show improved predictive performance, consistent with the grater statistical stability of graph properties in larger dislocation configurations their reduced sensitivity to finite size fluctuations. Consequently, the relationship between network topology and stress becomes more consistently represented in the graph properties, allowing the regression model to identify the underlying dependence reliably.
The ranking of representation performance is largely preserved between the two size regimes indicates that this is size effect rather than as artifact of any single representation.

A notable feature of these results is their consistency across the five graph representations $R_i$.
Although $R_0$-$R_4$ abstract the same dislocation configuration into an abstract graph in very different ways, the combination of graph properties and strain rate predicts the yield stress to similar high accuracy in every case, with $R^2$ varying only modestly between representations.
This robustness suggests that the predictive relationship is not tied to any particular graph construction how the network is encoded.
Its topology together with the loading rate is sufficient to capture the yield stress.
In other words, the graph-theoretic properties of the dislocation network carries relevant information that is preserved across representations and can be mapped to mechanical properties rather than a feature of an individual graph construction.

Finally, the results are consistent with the low intrinsic dimensionality of the graph property set revealed by the principal component analysis.
The seven graph properties collapse to roughly three groups for different representations, and, correspondingly, a five-property subset per representation is sufficient to saturate predictive accuracy.
The full seven properties set is not required.
Taken together, the random forest results demonstrate that a compact set of graph properties of the dislocation network combined with the loading rate is sufficient to predict the yield stress to within a few per cent of variance, and that this holds robustly across graph representations, across sizes and orientations.
The next step is if the predictive power of graph properties holds along a full strain path.

\section{Conclusion}\label{sec:conclusion}

We present an evaluation of dislocation networks from discrete dislocation dynamics represented a abstract graphs using five distinct graph representations ($R_0$-$R_4$), ranging from its edge-level network to loop connectivity, junction organization, and the connectivity of extended chains.
All representations preserve different aspects of the original dislocation network and therefore provide complementary structural description.
In particular, the reduced representations simplify the network by collapsing extended chains while retaining selected features of its junction structure, enabling more efficient characterization of network topology.
Comparing graph properties across all representations provides a means of assessing how the choice of structural representation influences the information available for predicting mechanical responses.
More broadly, our framework offers a systematic approach to evaluating the predictive power of dislocation network topology.
Each dislocation network is characterized using a compact set of graph properties.
Using a random forest regression model with strain rate-stratified cross-validation, we demonstrate that these graph properties, together with applied strain rate, predict yield stress with high accuracy.
For the larger networks, the model achieves $R^2$ values of between $0.86$-$0.94$ at yield point.
Among the five graph representations, $R_1$ exhibits slightly improved predictive performance than the others, as shown in Figure~\ref{fig:random_forest_3&4_size}.

Three points emerge from these results.
First, the predictive accuracy is consistent across all five representations, indicating that the link between network topology and mechanical response is a robust feature of the graph properties rather than an artifact of any single way of constructing a graph of a dislocation network.
Second, the accuracy improves systematically for larger sizes, as expected when finite size fluctuations in the topological properties diminish.
In other words, network measures make more sense when there is an actual network to characterize.
Third, principal component analysis shows that the seven graph properties span only about three effective groups because some graph properties are (anti)correlated for dislocation networks.
Consequently, a smaller subset of graph properties is sufficient to capture the topology-yield stress relationship without loss of predictive power.

Our findings demonstrate that graph properties provide a compact description of dislocation networks, capturing sufficient structural information to predict yield stress across different network representations and system sizes.
The good predictive performance of the random forest model for larger systems suggests that graph properties capture structural features that remain relevant as the network size increase, potentially reflecting aspects of \textit{collective} dislocation behavior.
These findings open opportunities to investigate how graph properties can contribute to continuum descriptions of the dislocation plasticity, either as state variables or by providing structural descriptors that bridge discrete dislocation networks and continuum-scale models.
Extending this framework to larger systems, more diverse network configurations, and flow stress predictions could further clarify the relationship between network topology, collective dislocation dynamics, and macroscopic mechanical response.
Ultimately, applying the same approach to experimentally-reconstructed dislocation structures may help establish graph-based characterization as a link between dislocation microstructure and continuum descriptions of plastic deformations at an abstract mesoscale.

\section*{Acknowledgements}

\section*{Declarations}
\begin{itemize}
	\item Funding\\
	G.L. and M.S. gratefully acknowledge support by the European Union by ERC grant DISCO-DATA, Project No. 101161287. The views and opinions expressed are, however, those of the authors only and do not necessarily reflect those of the European Union or the European Research Council Executive Agency. Neither the European Union nor the granting authority can be held responsible for them.
	\item Conflict of interest/Competing interests\\
	Not applicable
	\item Data availability\\
	\href{https://doi.org/10.5281/zenodo.24006153}{ZENODO}
	\item Code availability\\
	\href{https://gitlab.ruhr-uni-bochum.de/icams-mids/dislocation_graps.git}{GitLab}
	\item Author contribution\\
	\textbf{G.L.}: Methodology, Software, Validation, Formal analysis, Investigation, Data Curation, Writing - Original Draft, Writing - Review \& Editing, Visualization. \textbf{M.S.}: Conceptualization, Methodology, Formal analysis, Resources, Supervision, Writing - Review \& Editing, Project administration.
\end{itemize}

\begin{appendices}

%




\end{appendices}

\bibliography{sn-bibliography}

\end{document}